\documentclass[11pt]{article}
\usepackage[utf8]{inputenc}
\usepackage[margin=1in]{geometry}
\usepackage{graphicx}
\usepackage{hyperref}
\usepackage{amsmath}
\usepackage{amssymb}

\title{Vibe Coding: Toward an AI‑Native Paradigm for Semantic and Intent‑Driven Programming}
\author{Vinay Bamil\\
\texttt{vbamil@my.ggu.edu}\\
United States}
\date{}

\begin{document}

\maketitle

\begin{abstract}
Recent advances in large language models have enabled developers to generate
software by conversing with artificial intelligence systems rather than writing
code directly.  This paper introduces \emph{vibe coding}, an emerging
AI‑native programming paradigm in which a developer specifies high‑level
functional intent alongside qualitative descriptors of the desired ``vibe'' of
the solution (tone, style, or emotional resonance), and an intelligent agent
transforms those specifications into executable software.  We formalize the
definition of vibe coding, propose a reference architecture comprising an
intent parser, a semantic embedding engine, an agentic code generator, and an
interactive feedback loop, and present a hypothetical implementation.  We
compare vibe coding with declarative, functional, and prompt‑based
programming and discuss its implications for software engineering,
human–AI collaboration, and responsible AI practice.  While early reports
highlight substantial productivity gains and democratization of development
[\cite{premkumar2025,waters2025}], recent studies caution that
AI‑generated code often contains vulnerabilities and may even slow
experienced developers [\cite{metr2025,culafi2025,wired2025}].  We
examine these benefits and risks, identify key challenges such as
alignment, reproducibility, bias, explainability, maintainability, and
security, and outline future directions and open research questions,
including AI‑native
development environments, vibe‑tuned models, multi‑agent collaboration,
and responsible governance.
\end{abstract}

\section{Introduction}
Programming abstractions have steadily increased the expressiveness of
software development, moving from machine instructions to high‑level
languages and paradigms that hide implementation detail.  Recent progress in
large language models (LLMs) and AI coding assistants has enabled a
new mode of development where programmers describe desired behavior in natural
language and models generate code that satisfies those requirements.
Inspired by this trend, Andrej Karpathy coined the term \emph{vibe coding}
in February 2025 to describe a form of coding in which one ``fully gives in to
the vibes''—issuing high‑level voice or text commands to a coding agent and
accepting its suggestions without carefully reading the diffs
[\cite{premkumar2025,waters2025,evans2025,willison2025a}].  Popular
media and blog posts quickly picked up the concept; by March 2025,
the Merriam‑Webster online dictionary listed ``vibe coding'' as a slang term
for ``writing computer code in a somewhat careless fashion, with AI
assistance'' [\cite{waters2025}].

Advocates argue that vibe coding allows developers to focus on the big
picture—what they want to build and how it should feel—while the AI handles
implementation detail.  Early adopters claim that projects built with
vibe coding can be completed in a fraction of the time required by
traditional methods and that the approach lowers the barrier to entry for
people with little programming experience [\cite{premkumar2025,waters2025}].
Recent surveys reported that nearly 44\% of developers were already
using AI coding tools by 2023 and that projects using vibe coding can see
productivity gains of up to 55\% [\cite{premkumar2025}].  Nevertheless,
vibe coding has also attracted criticism.  Commentators note that the
practice often involves accepting AI‑generated code without understanding
its internal logic [\cite{kreutzbender2025,willison2025a}] and that
inadequate review can lead to software that is insecure or
unpredictable [\cite{culafi2025,wired2025}].  A randomized controlled
trial found that experienced open‑source developers using early‑2025 AI
coding tools actually took 19\% longer to complete tasks, contrary to
their expectation of a speedup [\cite{metr2025}].

This paper explores vibe coding as a distinct programming paradigm.
We define vibe coding formally, situate it within the landscape of
programming paradigms, and analyze its architecture and implementation.
Throughout the paper we adopt a critical lens, drawing on the growing
literature to highlight both the promise and pitfalls of this
AI‑native approach.  We argue that while vibe coding has the potential to
democratize software development and accelerate prototyping, realizing
its benefits responsibly will require careful attention to alignment,
reproducibility, security, and governance.

\section{Background}

\subsection{Evolution of Programming Paradigms}
Traditional paradigms such as imperative and object‑oriented programming
require developers to specify step‑by‑step instructions.  Declarative
languages allow programmers to describe what the program should accomplish
without prescribing how; examples include SQL for data queries and
HTML for document structure.  Functional programming emphasizes pure
functions and immutable data, aiming to make programs easier to reason
about.  Each of these paradigms attempts to raise the level of abstraction
and reduce the gap between the programmer's intent and the resulting code.

The rise of machine learning has introduced \emph{prompt‑based
programming}, in which developers provide natural language prompts and
LLMs generate code to satisfy those prompts.  Tools like GitHub
Copilot and Amazon CodeWhisperer pioneered this approach, enabling developers
to request code snippets or unit tests via descriptive comments.  Prompt
engineering has become a discipline in its own right, with practitioners
learning to craft effective instructions that guide LLMs toward desired
behaviors [\cite{premkumar2025}].

\subsection{From Prompt Programming to Vibe Coding}
Vibe coding builds on prompt programming but goes further in two ways.
First, it places equal emphasis on the \emph{functional intent} of the
software and the \emph{vibe}—the emotional tone, style, or user
experience the developer envisions.  In vibe coding the prompt might
include descriptors like ``playful and engaging'' or ``minimalist and
professional,'' and the agentic generator uses these descriptors to
select language, colors, or architectural patterns accordingly
[\cite{premkumar2025,waters2025}].  Second, vibe coding often entails
full trust in the AI's output: proponents describe copying and pasting
code without reading the diffs, adjusting parameters by asking the
agent to ``decrease the padding on the sidebar'' or ``add a mascot'' until
the result looks right [\cite{willison2025a}].

Simon Willison has distinguished vibe coding from general AI‑assisted
programming.  He notes that when a developer uses an LLM but carefully
reviews and understands every generated line of code, they are not vibe
coding; vibe coding, in his view, entails building software without
reviewing the code that the model writes [\cite{willison2025a}].  This
distinction has led to the emergence of \emph{vibe engineering}, a term
coined by Willison in October 2025 to describe a more disciplined use of
AI coding agents.  Vibe engineering encourages experienced engineers to
combine LLMs with best practices such as automated testing, planning,
documentation, and code review to produce maintainable,
production‑quality software [\cite{willison2025b}].  The ongoing debate
between vibe coding and vibe engineering underscores the need for clear
terminology and responsible practices.

\begin{table}[ht]
\centering
\caption{Comparison of programming paradigms}
\label{tab:paradigms}
\begin{tabular}{l p{3.2cm} p{3.2cm} p{4cm}}
\hline
Paradigm & Primary emphasis & Typical examples & Role of human developer \\
\hline
Imperative & Detailed step-by-step instructions & C, Java, Python & Writes and maintains code at syntactic level \\
Declarative & Describes what needs to be achieved & SQL, HTML, Haskell & Specifies goals and constraints; runtime derives how \\
Prompt-based & Natural-language specifications & GitHub Copilot, Amazon CodeWhisperer & Writes prompts and integrates AI-generated snippets \\
Vibe coding & Intent, tone and context descriptors & Emerging AI-native coding assistants & Specifies functional intent and vibe; collaborates with autonomous agent \\
\hline
\end{tabular}
\end{table}

\section{Definition of Vibe Coding}
We formally define \textbf{vibe coding} as an AI‑assisted software
development approach in which the programmer communicates their
\emph{functional intent}, \emph{emotional tone or style (vibe)}, and
\emph{contextual constraints} to an intelligent agent, which then
generates source code and artifacts to satisfy those specifications.  Let
$I$ denote the functional requirements, $E$ denote the vibe descriptors,
and $C$ denote any additional context or constraints (platform,
performance requirements, external APIs, etc.).  Vibe coding can be
conceptualized as a mapping
\[
\mathcal{A} : (I,E,C) \;\longmapsto\; P\,,
\]
where $\mathcal{A}$ is an AI coding agent and $P$ is the resulting
program.  The agent is responsible for selecting algorithms, data
structures, user interface elements, and textual wording that align with
$E$ while fulfilling $I$ under $C$.

A crucial aspect of vibe coding is the shift in the human developer's
role.  The developer focuses on specifying \emph{what} the software
should do and \emph{how it should feel}, not on writing or verifying the
underlying code.  As Willison remarks, if you review and understand
every line that the LLM writes, you are not vibe coding but using the
model as a typing assistant [\cite{kreutzbender2025,willison2025a}].
True vibe coding implies trust in the agent to handle low‑level details
and an iterative loop in which the developer evaluates the behavior and
requests high‑level adjustments rather than debugging code.

\section{Architecture}
Implementing vibe coding in practice requires a system that can capture
natural language specifications, interpret vibe descriptors, generate
code, and iterate based on feedback.  We propose an architecture
consisting of four primary components, illustrated conceptually in
Figure~\ref{fig:architecture}.  A textual description of the diagram
appears within the figure caption for clarity.

\begin{figure}[t]
  \centering
  \includegraphics[width=0.9\textwidth]{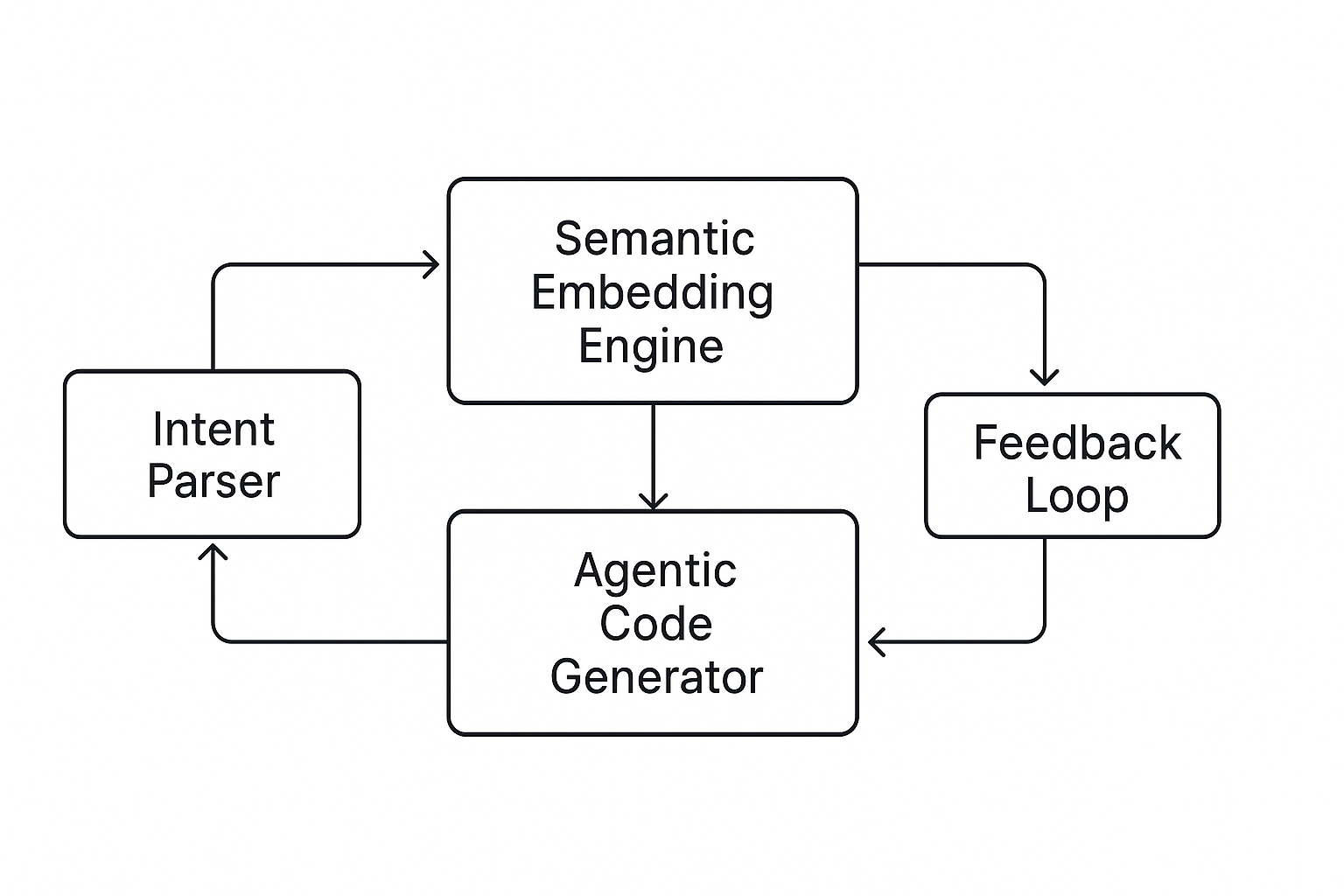}
  \caption{Illustrative architecture diagram for vibe coding.  The
  diagram depicts four interacting components: an \emph{Intent Parser}, a
  \emph{Semantic Embedding Engine}, an \emph{Agentic Code Generator}, and
  a \emph{Feedback Loop}, connected sequentially by arrows.}
  \label{fig:arch_diagram_img}
\end{figure}

\begin{figure}[t]
  \centering
  \fbox{\begin{minipage}{0.9\textwidth}
  \small
  \textbf{Architecture diagram description.}  The developer writes a natural
  language prompt that describes the desired functionality and vibe.  The
  \emph{Intent Parser} processes this prompt, extracting functional
  requirements, vibe descriptors, and constraints.  The parsed output is
  passed to a \emph{Semantic Embedding Engine}, which encodes the intent and
  vibe into vectors and retrieves relevant code patterns or examples.  These
  embeddings guide an \emph{Agentic Code Generator}—typically a large language
  model with planning and self‑reflection capabilities—that iteratively
  generates code, executes it in a sandbox, and reasons about errors.  A
  \emph{Feedback Loop} evaluates the generated output through tests and
  human feedback; it supplies corrections or refinements to the earlier
  stages until the final program satisfies the specified intent and vibe.
  \end{minipage}}
  \caption{High‑level architecture of a vibe coding system.  Natural
  language prompts are transformed into executable code through a pipeline
  consisting of an intent parser, semantic embedding engine, agentic
  code generator, and feedback loop.  Human input provides high‑level
  guidance and evaluations throughout the process.}
  \label{fig:architecture}
\end{figure}

\subsection{Intent Parser}
The first stage converts the developer’s free‑form description into a
structured representation.  Given a prompt like ``Create a personal finance
tracker that feels approachable and fun, with bright colors, friendly
messages, and offline storage,'' the intent parser identifies the core
functional goal (finance tracking with budgeting features), vibe descriptors
(approachable, fun, bright colors, friendly messages), and constraints
(mobile platform, offline capability).  Ambiguities are resolved either by
clarification prompts or by adopting reasonable defaults.  The parser may
produce an intermediate data structure (e.g., JSON or a set of
logical propositions) summarizing the extracted information.

\subsection{Semantic Embedding Engine}
To steer a generative model, the high‑level description must be translated into
representations that the agent can leverage.  The semantic embedding engine
encodes intent and vibe using vector embeddings.  For example, ``playful and
cartoon‑like'' may be mapped to a vector proximate to UI patterns known to
invoke joy.  The engine can also retrieve examples or templates matching
the vibe, such as CSS snippets for bright color palettes or code patterns
for gamified interfaces.  These embeddings and examples are provided to
the agentic generator as guidance.  The engine may incorporate context such
as organizational style guides or pre‑existing components, ensuring that
generated code integrates with existing systems.

\subsection{Agentic Code Generation}
At the heart of the system is an agentic code generator—an LLM capable of
planning, reasoning, and acting.  Rather than predicting code token by
token in a single pass, the agent can outline a plan, decompose tasks
into subtasks, generate code modules, execute them in a sandbox, and
self‑correct based on runtime feedback.  Recent models such as
Claude Code, Gemini CLI, and Codex CLI provide agentic capabilities and
support iterative refinement [\cite{willison2025b}].  The agent receives
the embedded intent and vibe as context, along with any relevant code
templates, and generates a candidate program.  It may produce comments or
pseudo‑code before writing full implementations, and it can call external
tools to compile, test, or run the code.  When errors occur or tests
fail, the agent uses the feedback to adjust the code without human
intervention.  The generator stops once the program passes tests and the
developer is satisfied with the vibe.

\subsection{Feedback Loop}
Vibe coding is inherently interactive.  After code generation, the system
evaluates whether the output meets both functional requirements and vibe
expectations.  Evaluation mechanisms include automated tests, static
analysis, security scans, and subjective assessments.  For instance, if a
finance tracker’s UI messages sound overly formal rather than playful, the
developer can request more enthusiasm or specify the use of emojis.  Dark
Reading’s poll of developers found that 24\% of respondents used vibe
coding tools with some success, while 41\% avoided them due to security
risks [\cite{culafi2025}].  The feedback loop allows developers to
iteratively refine the output, addressing bugs, adjusting tone, and
ensuring compliance with constraints.  Maintaining state across iterations
is essential; the agent must remember previous decisions and incorporate
corrections coherently.  When integrated into an IDE, the feedback loop
resembles a conversation with the AI assistant, enabling rapid cycles of
improvement.

\section{Implementation}
To illustrate the paradigm, consider a scenario where a developer wants to
build a to‑do list web application aimed at children.  They specify the
following prompt: ``Create a to‑do list app for kids.  It should have a fun,
cartoon‑like interface with bright colors and playful icons.  When a task is
completed, congratulate the child with an encouraging message.  The app must
run offline and store data in the browser.''  The system processes this
prompt through the stages above.  The intent parser extracts the goal
(to‑do list with offline storage), vibe (fun, cartoon‑like, bright colors,
playful icons, encouraging messages), and constraint (offline operation).
The embedding engine retrieves sample CSS for cheerful color schemes and
encodes ``encouraging'' to guide wording.  The agentic generator, using an
LLM with agentic scaffolding, plans the application: it decides to use
HTML/CSS/JavaScript, localStorage for offline data, and a simple event‑driven
UI.  It produces code for an input field, a list rendering function, and
event handlers for adding and completing tasks.  For the vibe, it chooses a
 bright palette, uses colourful icons, and writes messages like ``Great job! Keep it up!''—optionally accompanied by a simple thumbs up symbol—when tasks are checked off.

After the first iteration, the developer runs the app.  Suppose the
functionality works but the vibe feels too generic.  The developer enters
feedback: ``Make the tone more playful.  Add a cartoon mascot that cheers
when all tasks are done, and use more emojis in messages.''  The system
updates the vibe descriptors and regenerates the UI, adding a mascot image
that appears on task completion and sprinkling additional emoji in
notifications.  Each iteration takes seconds, allowing rapid convergence on
a kid‑friendly application.  This example illustrates how vibe coding
encourages developers to iterate on intent and style rather than
hand‑crafting code.

\section{Evaluation}
Assessing vibe coding requires metrics beyond conventional software
quality.  Functional correctness remains necessary: automated tests and
acceptance criteria must verify that the generated program meets its
specifications.  However, vibe coding introduces qualitative dimensions.  We
propose four categories of evaluation:

\begin{table}[ht]
\centering
\caption{Evaluation dimensions for vibe coding}
\label{tab:evaluation}
\begin{tabular}{l p{7.5cm}}
\hline
Dimension & Description \\
\hline
Functional metrics & Correctness, performance, resource usage, and efficiency compared to requirements and benchmarks \\
Vibe alignment & Degree to which generated software matches intended tone, style, and user experience \\
Security and safety & Presence of vulnerabilities, adherence to secure coding practices, and absence of hallucinated functionality \\
Developer experience & Usability, cognitive load, and perceived productivity for developers interacting with the system \\
\hline
\end{tabular}
\end{table}

\paragraph{Functional metrics.}  Standard metrics apply: coverage of
requirements, performance, resource consumption, and error rates.  LLM
generators can achieve high accuracy on benchmark suites and often
incorporate optimized algorithms [\cite{premkumar2025}].  Yet recent
randomized controlled trials found that experienced developers using
AI coding tools sometimes perform tasks more slowly; the METR study
reported that developers took 19\% longer to complete issues when allowed
to use early‑2025 AI tools, despite expecting a 24\% speedup
[\cite{metr2025}].  Such findings underscore that functional efficiency
depends on user experience and tool maturity.

\paragraph{Vibe alignment.}  Capturing whether the software matches the
intended tone or style is inherently subjective.  Human evaluation—user
studies where target users rate the software’s friendliness, professionalism,
or playfulness—provides one measure.  Automated proxies such as sentiment
analysis of UI text, diversity of color palettes, or readability scores can
offer objective signals.  For example, if a developer requests a playful
vibe, the system might analyze messages to ensure positive sentiment and
informal language.  Creating benchmark datasets with predefined vibes and
ground‑truth assessments remains an open research problem.

\paragraph{Security and safety.}  Security is a key concern.  AI‑generated
code frequently contains vulnerabilities, and LLMs are prone to hallucinate
nonexistent functions or misinterpret API usage.  Dark Reading notes that
while many organizations experiment with AI coding assistants, 41\% of
respondents avoided vibe coding due to security risks, and experts
warn that publishing an AI‑generated application without human review can be
a recipe for disaster [\cite{culafi2025}].  Wired reports that AI code
trained on old, vulnerable software can reintroduce past vulnerabilities
into new projects; different developers using the same model can produce
different outputs, complicating reproducibility and accountability
[\cite{wired2025}].  Security evaluation thus requires static analysis,
dynamic testing, and careful audit of external dependencies.  Policies and
guidelines—such as those adopted by governments for AI coding
assistants—should be integrated into the evaluation loop.

\paragraph{Developer experience.}  Vibe coding fundamentally changes the
developer’s workflow.  User studies can assess whether developers find
vibe coding intuitive, whether it reduces cognitive load, and whether
they perceive increased creativity or flow.  Surveys indicate that
developers appreciate the ability to quickly prototype ideas but remain
concerned about lack of control and transparency [\cite{culafi2025}].
Time‑and‑motion studies comparing vibe coding with traditional coding
across tasks of varying complexity can quantify changes in productivity and
stress.  The discrepancy between perceived speedup and actual slowdown
reported in the METR study suggests the importance of empirical
measurement [\cite{metr2025}].

In summary, evaluation should combine objective tests with subjective
assessments and incorporate security audits.  The field would benefit from
standardized benchmarks that include both functional requirements and
vibe descriptors.

\section{Challenges}
Vibe coding introduces numerous challenges that must be addressed to
transition the paradigm from toy projects to production systems.

\paragraph{Intention alignment and ambiguity.}  Translating vague or
subjective descriptions into code is difficult.  Phrases like ``edgy and
cool'' or ``approachable'' can be interpreted differently by developers and
models.  Misalignment between user intent and model interpretation can
lead to outputs that do not meet expectations.  Interactive clarification
and structured representations of vibe may help reduce ambiguity, but
achieving reliable alignment remains an open problem.

\paragraph{Reproducibility and consistency.}  Stochastic generative
models can produce different outputs for the same prompt across runs
[\cite{wired2025}].  This variability complicates version control and
collaboration.  Two developers may obtain divergent codebases for the same
specification, making merging and maintenance difficult.  Approaches such
as deterministic sampling or canonicalization of prompts may improve
consistency, but the broader challenge of reproducibility persists.

\paragraph{Bias and ethical concerns.}  The training data of LLMs contains
social biases, and vibe descriptors can inadvertently amplify them.  For
example, instructing an AI to produce a ``professional'' tone might result
in overly formal or exclusionary language.  Ethical guidelines and
content filtering must be incorporated into the generation and review
process.  Additionally, there is potential for misuse: vibe coding could
be exploited to automate phishing or produce persuasive content at scale.

\paragraph{Explainability and transparency.}  AI‑generated code can be a
black box.  Without clear rationale for design choices, it becomes
difficult to trust or debug the software.  Tools that provide natural
language explanations or rationales for generated code could improve
transparency.  However, such explanations are themselves generated by
models and may not reflect true reasoning.  Research into verifiable
explanations for code synthesis is needed.

\paragraph{Maintainability and debugging.}  Code produced via vibe coding
may not adhere to established architectural patterns.  Developers asked to
maintain or extend AI‑generated code often find it hard to understand,
especially if they did not participate in the generation process.  The
lack of comments, unusual naming conventions, or unconventional control
flow can impede maintenance.  Integrating software engineering best
practices—tests, documentation, modular design—into the agentic
workflow, as advocated by vibe engineering [\cite{willison2025b}], can
ameliorate this issue.

\paragraph{Security and safety.}  AI‑generated code frequently contains
critical vulnerabilities.  Dark Reading reports that developers and
security professionals worry about vulnerabilities and hallucinations in
AI‑generated code; only 24\% of surveyed respondents report using vibe
coding successfully, while 41\% avoid it due to security concerns
[\cite{culafi2025}].  Wired highlights that AI may reproduce outdated
open‑source vulnerabilities and that inconsistent outputs undermine
accountability [\cite{wired2025}].  The METR study found that developers
using AI tools sometimes overlook security practices because the agent
handles low‑level details [\cite{metr2025}].  To mitigate these risks,
security scans and human review must be integrated into the workflow.  In
high‑stake domains, vibe coding should be restricted or augmented with
formal verification.

\section{Future Directions and Open Research Questions}
Vibe coding is a nascent paradigm with many open research directions.  In
this section we suggest areas where further study could expand our
understanding of the paradigm.  These are not proposals for this
manuscript but rather recommendations for researchers and practitioners
who wish to advance the field.

\paragraph{AI‑native development environments.}  Future IDEs could be
designed around natural language interaction and vibe specification
instead of text editing.  Developers might sketch UI layouts and describe
desired feelings while the environment generates code and previews in real
time.  Multi‑modal interfaces (voice, touch, gesture) could enable more
expressive communication with coding agents.  Integrating version control,
testing, and deployment into such environments would streamline the
workflow.

\paragraph{Vibe‑tuned models and personalization.}  General‑purpose
models may struggle to capture specific tones or domain‑specific styles.
Training or fine‑tuning models on curated datasets corresponding to
particular vibes (e.g., playful UIs, formal business apps) could improve
alignment.  Personalization techniques like reinforcement learning from
human feedback could adapt agents to individual developers’ preferences.
Providing explicit ``vibe embeddings'' as control inputs may allow
continuous interpolation between styles.

\paragraph{Collaborative multi‑agent systems.}  Instead of a single
monolithic agent, future vibe coding frameworks may employ multiple
specialized agents.  One agent could generate code, another could test
security, a third could check vibe alignment, and a fourth could manage
dependencies.  Coordinating these agents in a pipeline or swarm could
improve reliability and reduce single‑agent failure modes.  Research into
communication protocols and task allocation among AI agents is needed.

\paragraph{Responsible AI and governance.}  As vibe coding becomes
prevalent, organizations must establish policies for safe usage.  This
includes guidelines on which tasks are appropriate for AI generation,
requirements for human review, standards for documentation and testing,
and processes for handling intellectual property.  Government guidelines
for AI coding assistants, such as those referenced in Dark Reading’s
coverage [\cite{culafi2025}], may inform best practices.  Educating
developers on prompt design, threat modeling, and AI limitations will be
essential.  Legal questions around ownership and liability for
AI‑generated code also warrant attention.

\section{Conclusion}
Vibe coding heralds a shift from manual, syntax‑driven programming to
semantic, intent‑driven collaboration with AI.  We have formalized the
concept, situating it among existing paradigms and detailing a reference
architecture that includes an intent parser, semantic embeddings, an
agentic generator, and an interactive feedback loop.  An illustrative
implementation demonstrates how developers can iteratively guide an AI
assistant to produce an application that embodies both functionality and
desired vibe.  Our evaluation highlights that while vibe coding promises
productivity and democratization of development, empirical studies reveal
mixed results: some developers experience slowdowns and security risks
[\cite{metr2025,culafi2025,wired2025}].  We have outlined challenges,
including alignment, reproducibility, bias, transparency, maintainability
    and security, and proposed future directions and open research
    questions such as AI‑native development environments, vibe‑tuned
    models, multi‑agent collaboration, and responsible governance.
    Ultimately, vibe coding may evolve into a
spectrum of practices ranging from experimental prototyping to disciplined
\emph{vibe engineering}.  Harnessing its potential responsibly will
require interdisciplinary collaboration between AI researchers, software
engineers, human–computer interaction experts, and ethicists to ensure
that the code we generate not only works, but truly resonates with human
values and intent.

\section*{Disclosure of Generative AI Assistance}
\label{sec:disclosure}
In accordance with arXiv’s policy on the use of generative AI language
tools, we disclose that large language models were used as a tool to
assist in drafting and organizing some sections of this article.  All
text generated by AI was reviewed, edited, and supplemented by the
author to ensure accuracy, coherence, and scholarly rigor.  No AI
system is credited as an author; the named author takes full
responsibility for the content and stands by the integrity of the
research presented herein.

\end{document}